\documentclass[aps,prl,twocolumn,showpacs,floatfix,nofootinbib,groupedaddress,superscriptaddress]{revtex4}

\usepackage{bm,bbm,graphics,graphicx,times,color,multirow}

\begin{document}

\title{Direct Characterization of Quantum Dynamics}
\author{M. Mohseni}
\affiliation{Department of Physics, University of Toronto, 60 St.
George St., Toronto, Ontario, M5S 1A7, Canada}
\affiliation{Department of Chemistry, University of Southern
California, Los Angeles, CA 90089, USA}
\author{D. A. Lidar}
\affiliation{Department of Chemistry, University of Southern
California, Los Angeles, CA 90089, USA} \affiliation{Department of
Electrical Engineering, and Department of Physics, University of
Southern California, Los Angeles, CA 90089, USA}

\begin{abstract}
The characterization of quantum dynamics is a fundamental and
central task in quantum mechanics. This task is typically addressed
by quantum process tomography (QPT). Here we present an alternative
``direct characterization of quantum dynamics'' (DCQD) algorithm. In
contrast to all known QPT methods, this algorithm relies on
error-detection techniques and does not require any quantum state
tomography. We illustrate that, by construction, the DCQD algorithm
can be applied to the task of obtaining \emph{partial information}
about quantum dynamics. Furthermore, we argue that the DCQD
algorithm is experimentally implementable in a variety of prominent
quantum information processing systems, and show how it can be
realized in photonic systems with present day technology.

\end{abstract}

 \maketitle

The characterization and identification of quantum systems are among
the central modern challenges of quantum physics, and play an
especially fundamental role in quantum information science
\cite{Nielsen:book}, and coherent control \cite{Rabitz:00}. A task
of general and crucial importance is the characterization of the
dynamics of a quantum system that has an unknown interaction with
its embedding environment. Knowledge of this dynamics is
indispensable, e.g., for verifying the performance of an
information-processing device, and for the design of
decoherence-mitigation methods. Such characterization of quantum
dynamics is possible via the method of standard quantum process
tomography (SQPT) \cite{SQPT,Nielsen:book}. SQPT consists of
preparing an ensemble of identical quantum systems in a member of a
set of quantum states, followed by a reconstruction of the dynamical
process via quantum state tomography. We identify three main issues
associated with the required physical resources in SQPT. (i) The
number of ensemble measurements grows exponentially with the number
of degrees of freedom of the system. (ii) Often it is not possible
to prepare the complete set of required quantum input states. (iii)
Information concerning the dynamical process is acquired
\emph{indirectly} via quantum state tomography, which results in an
inherent redundancy of physical resources associated with the
estimation of some superfluous parameters. To address (ii), the
method of ancilla-assisted process tomography (AAPT) was proposed
\cite{ArianoPRL01}. However, the number of measurements is the same
in SQPT and (separable) AAPT \cite{ArianoPRL01}.

\begin{figure}[tbp]
\includegraphics[width=7cm,height=1.4cm]{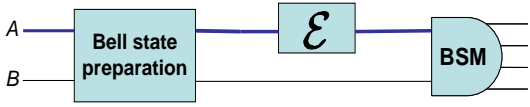}
\caption{Schematic of characterization algorithm for single qubit
case,
consisting of Bell state preparations, applying the unknown quantum map, $%
\mathcal{E}$, and Bell-state measurement (BSM).} \label{bsmf}
\end{figure}

Here we develop an algorithm for \emph{direct characterization of
quantum dynamics} (DCQD), which does not require quantum state
tomography. The primary system is initially entangled with an
ancillary system, before being subjected to the unknown dynamics.
Complete information about the dynamics is then obtained by
performing a certain set of error-detecting measurements. We
demonstrate that for characterizing a non-trace preserving quantum
dynamical map on $n$ qubits the number of required experimental
configurations is reduced from $2^{4n}$, for SQPT and
\emph{separable} AAPT, to $2^{2n}$ in DCQD. E.g., for a single
qubit, we show that one can fully characterize the quantum dynamics
by preparing one of four possible two-qubit entangled states, and a
Bell-state measurement (BSM) at the output. This is illustrated in
Fig.~\ref{bsmf} and Table~\ref{tabb}. We also discuss the
experimental feasibility of DCQD in a variety of quantum information
processing (QIP) systems, and show how it can be realized with
linear optics.

In principle, the number of required experimental configurations in
the AAPT scheme can be reduced by utilizing \emph{non-separable}
(global) measurements for the required state tomography, such as
mutually unbiased basis (MUB) measurements \cite{Wootters89}, or a
generalized measurement \cite{DArianoUQO02}. In general, these types
of measurements require \emph{many-body interactions} which are not
available experimentally. Here, we demonstrate that the DCQD
algorithm requires only $\mathcal{O}(n)$ single- and two-body
operations per experimental configuration.

We demonstrate the inherent applicability of the DCQD algorithm to
the task of partial characterization of quantum dynamics in terms of
coarse-grained quantities. Specifically, we demonstrate that for a
two-level system undergoing a sequence of amplitude and phase
damping processes, the relaxation time $T_{1}$ and dephasing time
$T_{2}$ can be simultaneously determined in one ensemble
measurement.

\textit{Quantum Dynamics}.--- The evolution of a quantum system
(open or closed) can, under natural assumptions, be expressed in
terms of a completely positive quantum dynamical map $\mathcal{E}$,
which can be represented as \cite{Nielsen:book}
\begin{equation}
{\mathcal{E}}(\rho )=\sum_{m,n=0}^{d^{2}-1}\chi _{mn}~E_{m}\rho
E_{n}^{\dagger }.
\end{equation}
Here $\rho $ is the system initial state, and the $\{E_{m}\}$ are a
set of (error) operator basis elements in the Hilbert-Schmidt space
of linear operators acting on the system, satisfying
$\text{Tr}(E_{i}^{\dagger }E_{j})=d\delta _{ij}$. The $\{\chi
_{mn}\}$ are the matrix elements of the superoperator $\bm{\chi}$,
which encodes all the information about the dynamics, relative to
the basis set $\{E_{m}\}$ \cite{Nielsen:book}. For an $n$ qubit
system, the number of independent matrix elements in $\bm{\chi}$ is
$2^{4n}$ for a non-trace-preserving map, or $2^{4n}-2^{2n}$ for a
trace-preserving map. The matrix $\bm{\chi}$ is positive Hermitian, and $%
\mathrm{Tr}\bm{\chi}\leq 1$. Thus $\bm{\chi}$ can be thought of as a
density matrix in Hilbert-Schmidt space, whence we often refer,
below, to its diagonal and off-diagonal elements as
\textquotedblleft quantum dynamical population\textquotedblright\
and \textquotedblleft coherence\textquotedblright , respectively.

\textit{Characterization of Quantum Dynamical Population}.--- Here
we demonstrate how to determine $\{\chi _{mm}\}$ in a single
(ensemble) measurement for a single qubit. Let us maximally entangle
two qubits $A$ (primary)\ and $B$
(ancillary) as $|\psi \rangle =(|0_{A}0_{B}\rangle +|1_{A}1_{B}\rangle )/%
\sqrt{2}$, and then subject only qubit $A$ to a map $\mathcal{E}$.
From here on, for simplicity, we denote $\mathcal{E\otimes I}$ by
$\mathcal{E}.$ In this case the error basis $\{E_{m}\}_{m=0}^{3}$
becomes the identity operator and the Pauli operators:
$\{I,X,Y,Z\}$. The state $|\psi \rangle $
is a $+1$ eigenstate of the commuting operators $Z^{A}Z^{B}$ and $X^{A}X^{B}$%
, i.e., it is stabilized under the action of these \textquotedblleft
stabilizer operators\textquotedblright , and it is referred to as a
\textquotedblleft stabilizer state\textquotedblright\
\cite{Nielsen:book}. Any non-trivial operator acting on the state of
the first qubit anticommutes with at least one of $Z^{A}Z^{B}$ and
$X^{A}X^{B}$, and therefore by measuring these operators we can
detect an arbitrary error on the first qubit, \textit{i.e.,} by
finding the eigenvalues of either one or both operators to be $-1$.
Measuring the observables $Z^{A}Z^{B}$ and $X^{A}X^{B}$ is
equivalent to a BSM, and can be represented by the four projection
operators $P_{0}=|\phi ^{+}\rangle \langle \phi ^{+}|$, $P_{1}=|\psi
^{+}\rangle \langle \psi ^{+}|$, $P_{2}=|\psi ^{-}\rangle \langle \psi ^{-}|$%
, $P_{3}=|\phi ^{-}\rangle \langle \phi ^{-}|$, where $|\phi ^{\pm
}\rangle =(|00\rangle \pm |11\rangle )/\sqrt{2}$, $|\psi ^{\pm
}\rangle =(|10\rangle \pm |01\rangle )/\sqrt{2}$, form the Bell
basis of the two-qubit system. The probabilities of obtaining the no
error outcome $I$, bit flip error $X^{A}$, phase flip error $Z^{A}$,
and both phase flip and bit flip errors $Y^{A}$, on the first qubit,
become $p_{m}=\mathrm{Tr}[P_{m}\mathcal{E}(\rho )]=\chi _{mm}$, for
$m=0,1,2,3$ respectively. Therefore, we can determine the quantum
dynamical population, $\{\chi _{mm}\}_{m=0}^{3}$, \emph{in a single
ensemble measurement} (e.g., by simultaneously measuring the operators $%
Z^{A}Z^{B}$ and $X^{A}X^{B}$) on multiple copies of the state $|\psi
\rangle $.

\textit{Characterization of Quantum Dynamical Coherence}.--- In
order to preserve the coherence ($\chi _{m\neq n}$) in the quantum
dynamical process, we perform a set of measurements such that we
always obtain partial information about the nature of the errors.
The simplest case is a measurement of the projection operators
$P_{+1}=|\phi ^{+}\rangle \langle \phi ^{+}|+|\phi ^{-}\rangle
\langle \phi ^{-}|$ and $P_{-1}=|\psi ^{+}\rangle \langle \psi
^{+}|+|\psi ^{-}\rangle \langle \psi ^{-}|$
corresponding to the eigenvalues $+1$ and $-1$ of measuring the stabilizer $%
Z^{A}Z^{B}$. The outcomes of this measurement represent the
probabilities of no bit flip error and bit flip error on qubit $A$,
without telling us anything about a phase flip error. Therefore, we
preserve only the coherence between operators $I$ and $Z^{A}$, and
also between the $X^{A}$ and $Y^{A}$ (which are represented by the
off-diagonal elements $\chi _{03}$ and $\chi _{12}$, respectively).
The required input state is a non-maximally entangled state $|\phi
_{C}\rangle =\alpha |00\rangle +\beta |11\rangle $, with $|\alpha
|\neq |\beta |\neq 0$, whose sole stabilizer is $Z^{A}Z^{B}$. Now,
by separating real and imaginary parts, the probabilities of no bit
flip error and bit flip error events become:
$\mathrm{Tr}[P_{+1}\mathcal{E}(\rho
)]=\chi _{00}+\chi _{33}+2\mathrm{Re}(\chi _{03})\langle Z^{A}\rangle $ and $%
\mathrm{Tr}[P_{-1}\mathcal{E}(\rho )]=\chi _{11}+\chi _{22}+2\mathrm{Im}%
(\chi _{12})\langle Z^{A}\rangle $, where $\langle Z^{A}\rangle
\equiv \mathrm{Tr}(\rho Z^{A})\neq 0$ (because $|\alpha |\neq |\beta
|\neq 0$),
with $\rho =|\phi _{C}\rangle \langle \phi _{C}|$. We already know the $%
\{\chi _{mm}\}$ from the population measurement described above, so
we can determine $\mathrm{Re}(\chi _{03})$ and $\mathrm{Im}(\chi
_{12})$. After measuring $Z^{A}Z^{B}$ the system is in either of the
states $\rho _{\pm
1}=P_{\pm 1}\mathcal{E}(\rho )P_{\pm 1}/\mathrm{Tr}[P_{\pm 1}\mathcal{E}%
(\rho )]$. Now we measure the expectation value of a (commuting)
normalizer operator $U$ (such as $%
X^{A}X^{B}$) \cite{normalizer}. Thus we obtain $\mathrm{Tr}[U\rho
_{+1}]=[(\chi_{00}-\chi_{33})\langle U\rangle +2i\mathrm{Im}(\chi _{03})\langle Z^{A}U\rangle ]/%
\mathrm{Tr}[P_{+1}\mathcal{E}(\rho )]$ and $\mathrm{Tr}[U\rho
_{-1}]=[(\chi _{11}-\chi _{22})\langle U\rangle -2i\mathrm{Re}(\chi
_{12})\langle Z^{A}U\rangle ]/\mathrm{Tr}[P_{-1}\mathcal{E}(\rho
)]$, where $\langle Z^{A}\rangle $, $\langle U\rangle $, and
$\langle Z^{A}U\rangle $ are all non-zero and already known.
Therefore, we can obtain the four independent real parameters needed
to calculate the coherence components $\chi _{03}$
and $\chi _{12}$, by simultaneously measuring, e.g., $Z^{A}Z^{B}$ and $%
X^{A}X^{B}$.

\begingroup\squeezetable
\begin{table}[bp]
\begin{ruledtabular}
\caption{One possible set of input states and measurements for
direct characterization of quantum dynamics ($\chi_{ij}$) for a
single qubit, where $|\alpha|\neq |\beta|\neq 0$, and
$\{|0\rangle,|1\rangle \}$, $\{|\pm\rangle \}$, $\{|\pm i\rangle\}$
are eigenstates of the Pauli operators $Z$, $X$, and $Y$.}
\begin{tabular}{c|cc|c}
\multicolumn{1}{c|}{input state}&\multicolumn{2}{c|}{Measurement}&\multicolumn{1}{c}{output}\\
 & Stabilizer & Normalizer & \\
 \colrule
$(|0\rangle|0\rangle+|1\rangle|1\rangle)/\sqrt{2}$& $Z^AZ^B,X^AX^B$ & N/A & $\chi_{00},\chi_{11},\chi_{22},\chi_{33}$\\
$\alpha|0\rangle|0\rangle+\beta|1\rangle|1\rangle$ & $Z^AZ^B$ & $X^AX^B$ & $\chi_{03},\chi_{12}$\\
$\alpha|+\rangle|0\rangle+\beta|-\rangle|1\rangle$ & $X^A Z^B$ & $Z^A X^B$ & $\chi_{01},\chi_{23}$\\
$\alpha|+i\rangle|0\rangle+\beta|-i\rangle|1\rangle$ & $Y^A Z^B$ & $Z^A X^B$ & $\chi_{02},\chi_{13}$\\
\end{tabular}\label{tabb}
\end{ruledtabular}
\end{table}
\endgroup

\begingroup\squeezetable
\begin{table}[tp]
\begin{ruledtabular}
\caption{Comparison of the required physical resources for
characterizing a non-trace preserving CP quantum dynamical map on
$n$ qubits. $N_{\text{in}}$ and $N_{\text{m}}$ respectively denote
the number of required input states and the number of
\textit{noncommutative} measurements for each input state, and
$N_{\text{exp}}\equiv N_{\text{in}}N_{\text{m}}$ is the number of
required experimental configurations.}

\begin{tabular}{lccccccccc}
 Scheme & $\text{dim}({\mathcal H})$ & $N_{\text{in}}$ & $N_{\text{m}}$&
 $N_{\text{exp}}$\\
\colrule
SQPT & $2^n$ & $4^n$ & $4^n$ & $16^n$\\
non-separable AAPT & $2^{2n}$ & 1 & $4^n+1$ & $4^n+1$\\
DCQD & $2^{2n}$ & $4^n$ & 1 & $4^n$ \\
\end{tabular}\label{tab}
\end{ruledtabular}
\end{table}
\endgroup

In order to characterize the remaining coherence elements of
$\bm{\chi}$ we make an appropriate change of basis in the
preparation of the two-qubit system. For characterizing $\chi _{01}$
and $\chi _{23}$, we can perform a Hadamard transformation on the
first qubit, as $H^{A}|\phi _{C}\rangle =\alpha |+\rangle |0\rangle
+\beta |-\rangle |1\rangle $, where $|\pm \rangle =(|0\rangle \pm
|1\rangle )/\sqrt{2}$. We then measure the
stabilizer operator $X^{A}Z^{B}$, and a normalizer such as $%
Z^{A}X^{B}$. For characterizing $\chi _{02}$ and $\chi _{31}$, we
prepare the system in the stabilizer state $S^{A}H^{A}|\phi
_{C}\rangle =\alpha |+i\rangle |0\rangle +\beta |-i\rangle |1\rangle
$, and measure the stabilizer operator $Y^{A}Z^{B}$ and a normalizer
such as $Z^{A}X^{B} $, where $S$ is the single-qubit phase gate, and
$|\pm i\rangle =(|0\rangle \pm i|1\rangle )/\sqrt{2}$. Therefore, we
can completely characterize the quantum dynamical coherence with
three BSM's overall. We note that the bases $\{|0\rangle ,|1\rangle
\}$, $\{|\pm \rangle \}$ and $\{|\pm i\rangle \}$ are \emph{mutually unbiased}%
, i.e., the inner products of each pair of elements in these bases
have the same magnitude \cite{Wootters89}.

A summary of the scheme for the case of a single qubit is presented
in Fig.~\ref{bsmf} and Table.~\ref{tabb}. This table implies that
the required resources in DCQD are as follows: (a) preparation of a
maximally entangled state (for population characterization), (b)
preparation of three other (nonmaximally) entangled states (for
coherence characterization), and (c) a fixed Bell-state analyzer. We
remark that a generalized DCQD algorithm for qudits, with $d$ being
a power of a prime, is possible, and will be the subject of a future
publication~\cite{MohseniLidar06}.

An additional feature of DCQD is that all the required ensemble
measurements, for measuring the expectation values of the stabilizer
and normalizer operators, can also be performed in a temporal
sequence on the \emph{same} pair of qubits with only \emph{one}
Bell-state generation. This is because at the end of each
measurement, the output state is in fact in one of the four possible
Bell states, which can be utilized as an input stabilizer state.

For characterizing a quantum dynamical map on $n$ qubits we need to
perform a measurement corresponding to a tensor product of the
required measurements for single qubits. An important example
is a QIP unit with $n$ qubits which has a $2^{n}$-dimensional Hilbert space (%
$\mathcal{H}$). DCQD requires a total of $4^{n}$ experimental
configurations for a complete characterization of the dynamics. This
is a \emph{quadratic advantage} over SQPT and separable AAPT, which
require a total of $16^{n}$ experimental configurations. In general,
the required state tomography in AAPT could also be realized by
non-separable (global) quantum measurements. These measurements can
be performed either in the same Hilbert space, with $4^{n}+1$
measurements, e.g., using a MUB measurement \cite{Wootters89}, or in
a \emph{larger} Hilbert space, with a single generalized measurement
\cite{DArianoUQO02}. A comparison of the required physical resources
is presented in Table~\ref{tab}. A detailed resource cost analysis
comparing DCQD to other QPT methods will be reported in a future
publication \cite{MRL-CQPT06}.

\textit{Physical Realization}.--- For qubit systems, the resources
required in order to implement the DCQD algorithm are Bell state
preparation and measurement, and single qubit rotations. These tasks
play a central role in quantum information science, since they are
prerequisites for quantum
teleportation, quantum dense coding and quantum key distribution \cite%
{Nielsen:book}. Because of their importance, these tasks have been
studied extensively and successfully implemented in a variety of
different quantum
systems, e.g., nuclear magnetic resonance (NMR) \cite{NMRbellstates} and trapped ions \cite%
{WinelandBellstates}. Thus, the DCQD algorithm is already within
experimental feasibility of essentially all systems which have been
used to demonstrate QIP principles to date.

\begin{figure}[tp]
\includegraphics[width=8.5cm,height=3.5cm]{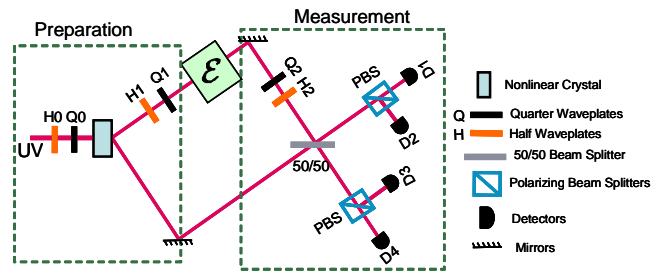}
\caption{A schematic layout of a linear-optical realization of the
DCQD algorithm. A pair of entangled photons, $%
(\left\vert H_{A}H_{B}\right\rangle +e^{i\protect\varphi }\left\vert
V_{A}V_{B}\right\rangle )/\protect\sqrt{2},$ is generated via
parametric
down-conversion in a nonlinear crystal from a UV pump laser, where $%
\left\vert H\right\rangle $ and $\left\vert V\right\rangle $
represent horizontal and vertical polarization \protect\cite%
{Kwiat96}. The quarter and half waveplates are used to create
non-maximally entangled states \protect\cite{White99}, and change
the preparation and measurement bases. Two-photon interferometry
occurs at the 50/50 beam splitters. This allows for discriminating
two of the four Bell states deterministically, by utilizing
polarizing beam splitters and photodetectors as analyzers. }
\label{scheme-fig}
\end{figure}

Here, we propose a specific linear-optical implementation,
realizable with present day technology -- see Fig.~\ref{scheme-fig}.
Using only linear-optical elements, at most 50\% efficiency in
discriminating among Bell states is possible \cite{Kwiat96}. The
number of ensemble preparations in this specific setup is thus
effectively increased by a factor of two over an implementation of
the DCQD algorithm using an ideal Bell state analyzer. However, even
in this optical realization, the number of experimental
configurations is still reduced by a factor of $2^{n}$ (for
characterizing a quantum dynamical process on $n$ qubits) over SQPT
and separable AAPT schemes. Alternatively, a small-scale and
deterministic linear-optical implementation of DCQD can in principle
be realized by a multirail representation of the qubits
\cite{small-scaleLO}.

To increase the efficiency of an all-optical Bell-state analyzer, we
have three different strategies: (i) introducing some nonlinearity,
such as an optical switch \cite{Kevin}, (ii) utilizing postselected
measurements, such as an optical \textsc{cnot} \cite{KLM}, which has
been demonstrated
experimentally for filtering Bell states with a fidelity of about 79\% \cite%
{BSFopticalCNOT}, (iii) employing hyperentanglement between pairs of
photons for a complete and deterministic linear-optical Bell-state
discrimination \cite{linearoptics BSM98,linearoptics BSM06}. The
latter method is based on the fact that the pairs of
polarization-entangled photons generated by parametric
down-conversion have intrinsic correlations in time-energy and
momentum. These additional degrees of freedom can be exploited in
order to distinguish between the subsets of Bell-states which
otherwise cannot be discriminated by a standard Hong-Ou-Mandel
interferometer. Note that employing the time-energy correlations in
the context of implementing the DCQD algorithm is feasible if the
quantum dynamical map acts only on the polarization degrees of
freedom.

\textit{Partial Characterization of Dynamics}.--- DCQD can be
applied, \emph{by construction}, to the task of partial
characterization of quantum dynamics, where we cannot afford or do
not need a full characterization of the system, or when we have some
\textit{a priori} knowledge about the dynamics. In particular, we
can substantially reduce the number of measurements, when we are
interested in estimating the coherence elements of the superoperator
for only specific \emph{subsets} of the operator basis and/or
subsystems of interest. E.g., we need to perform a single ensemble
measurement if we are required to identify only the coherence
elements $\chi _{03}$ and $\chi _{12}$ of a particular qubit. Here,
we present an example of such a task. Specifically, we demonstrate
that the DCQD algorithm enables the simultaneous determination of
coarse-grained physical quantities, such as the longitudinal
relaxation time $T_{1}$ and the transverse relaxation (or dephasing)
time $T_{2}$. Assume that we prepare a two-qubit system in the
non-maximally entangled state $|\phi _{C}\rangle =\alpha |00\rangle
+\beta |11\rangle $. Then we subject qubit $A$ to an amplitude
damping process for duration $t_{1}$, followed by a phase damping
process, for duration $t_{2}$. The elements of
the final density matrix then read $\langle 0|\rho _{f}|0\rangle =1-\exp (-%
\frac{t_{1}}{T_{1}})(1-|\alpha |^{2})$ and $\langle 0|\rho
_{f}|1\rangle
=\exp (-\frac{t^{\prime }}{2T_{2}^{\prime }})\alpha ^{\ast }\beta $, where $%
\frac{t^{\prime }}{T_{2}^{\prime }}=\frac{t_{2}}{T_{2}}+\frac{t_{1}}{T_{1}}$%
. In order to determine $T_{1}$, we measure the eigenvalues of the
stabilizer operator $Z^{A}Z^{B}$. We obtain either $+1$ or $-1$,
corresponding to the projective measurement $P_{+1}$, or $P_{-1}$.
The probabilities of either of these outcomes, e.g.,
$\mathrm{Tr}(P_{-1}\rho _{f})$
are related to $T_{1}$ through the relation $\frac{1}{T_{1}}=-\frac{1}{%
  t_{1}}\ln (1-\frac{2\mathrm{Tr}(P_{-1}\rho
  _{f})}{1-\mathrm{Tr}(Z^{A}\rho )}) $.
In order to obtain information about $T_{2}$, we measure the
expectation value of any normalizer of the input state, such as
  $X^{A}X^{B}$,
yielding $\frac{t^{\prime }}{T_{2}^{\prime }}=-2\ln \frac{\mathrm{Tr}%
(X^{A}X^{B}\rho _{f})}{\mathrm{Tr}(X^{A}X^{B}\rho )}$. Since the operators $%
Z^{A}Z^{B}$ and $X^{A}X^{B}$ commute, we can measure them
simultaneously. Therefore we can find both $T_{1}$ and $T_{2}$ in a
Bell-state measurement.

\textit{Outlook}.---One can combine the DCQD algorithm with the
method of maximum-likelihood estimation \cite{Kosut04}, in order to
minimize the statistical errors in each experimental configuration.
Moreover, a new scheme for \emph{continuous} characterization of
quantum dynamics can be introduced, by utilizing weak measurements
for the required quantum error detections in DCQD \cite{Ahn02}.

For quantum systems with controllable two-body interactions (e.g.,
trapped-ion and NMR systems), DCQD could have near-term experimental
applications for complete verification of \emph{small QIP units}.
For example, DCQD, reduces the number of required experimental
configurations for systems of $3$ or $4$ physical qubits from
$5\times 10^{3}$ and $6.5\times 10^{4}$ (in SQPT) to $64$ and $256$,
respectively. A similar scale-up can only be achieved by utilizing
non-separable AAPT methods. Complete characterization of such
dynamics would be essential for verification of quantum key
distribution procedures, teleportation units, quantum repeaters, and
more generally, in any situation in quantum physics where a few
qubits have a common local bath and interact with each other.
Furthermore, as demonstrated here, DCQD is inherently suited to
extract partial information about quantum dynamics. Several other
examples of such applications have been demonstrated. Specifically,
it has been shown that DCQD can be used for realization of
generalized quantum dense coding \cite{MohseniLidar06}. Moreover,
DCQD can be efficiently applied to (single- and two-qubit)
Hamiltonian identification tasks \cite{MRL-CQPT06}. Finally, the
general techniques developed here could be further utilized for
closed-loop control of open quantum systems.

This work was supported by NSERC (to M.M.), NSF CCF-0523675, ARO
W911NF-05-1-0440, and the Sloan Foundation (to D.A.L.). We thank R.
Adamson, J. Emerson, D.F.V. James, K. Khodjasteh, D.W. Leung, A.T.
Rezakhani, A.M. Steinberg, and M. Ziman for useful discussions.


\begin{thebibliography}{99}
\bibitem{normalizer} A normalizer operator is a unitary operator that preserves the stabilizer
subspace.
\end{thebibliography}
\end{document}